\documentclass[twocolumn, pra, superscriptaddress]{revtex4-1}
\usepackage{amssymb}
\usepackage{amsmath}
\usepackage{epsfig}
\usepackage{color}
\usepackage{graphics, graphicx}
\usepackage{bbold}
\usepackage{psfrag}
\usepackage{mathcomp}
\usepackage{subfigure}
\usepackage{verbatim}
\usepackage{color}
\usepackage[colorlinks, citecolor=blue]{hyperref}
\def\cp#1{\mathbf{#1}}

\begin{document}

\title{Stretching $p$-wave molecules by transverse confinements}
\author{Lihong Zhou}
\affiliation{Beijing National Laboratory for Condensed Matter Physics, Institute of Physics, Chinese Academy of Sciences, Beijing 100190, China}
\affiliation{University of Chinese Academy of Sciences, Beijing 100049, China}
\author{Xiaoling Cui}
\email{xlcui@iphy.ac.cn}
\affiliation{Beijing National Laboratory for Condensed Matter Physics, Institute of Physics, Chinese Academy of Sciences, Beijing 100190, China}
\date{\today}
\begin{abstract}
We revisit the confinement-induced $p$-wave resonance in quasi-one-dimensional(quasi-1D) atomic gases and study the induced molecules near resonance. 
We derive the reduced 1D interaction parameters and show that they can well predict the binding energy of shallow molecules in quasi-1D system. Importantly, these shallow molecules are found to be much more spatially extended compared to those in three-dimension (3D) without transverse confinement. Our results strongly indicate that a $p$-wave interacting atomic gas can be much more stable in quasi-1D near the induced $p$-wave resonance, where most weight of the molecule lies outside the short-range regime and thus the atom loss could be suppressed. 
\end{abstract}

\maketitle

{\it Introduction.} $P$-wave interaction is known to lead to intriguing quantum phenomena that are hardly accessible by pure $s$-wave interactions, such as the rich pairing mechanism in three-dimensions (3D) due to orbital degrees of freedom \cite{Gurarie, Yip}, the $p+ip$ topological superfluid in two-dimensions (2D) \cite{Read, Zoller} and the Majorana fermion in one-dimension (1D) from the classic Kitaev chain model \cite{Kitaev}. Given the highly tunable $p$-wave interaction via Feshbach resonance \cite{K40,K40_2,Toronto,Li6_1,Li6_2,Rb}, the ultracold atomic gases emerge as a promising platform for exploring these $p$-wave phenomena.
Nevertheless, the exploration of the $p$-wave effect in a 3D atomic gas has been largely impeded due to severe three-body losses near the usually narrow $p$-wave Feshbach resonance \cite{K40,K40_2,Toronto,Li6_1,Li6_2}. Recently, a number of open-channel-dominated broad $p$-wave resonances have been found in boson-boson mixtures \cite{Rb}, but still no clear evidence shows the losses can be well controlled \cite{private}. Because of the atom loss, so far the experiment can only probe the low-temperature physics of a resonant $p$-wave gas in the quasi-equilibrium regime within very short time scale \cite{Toronto}.

\begin{figure}[t]
\includegraphics[width=9cm]{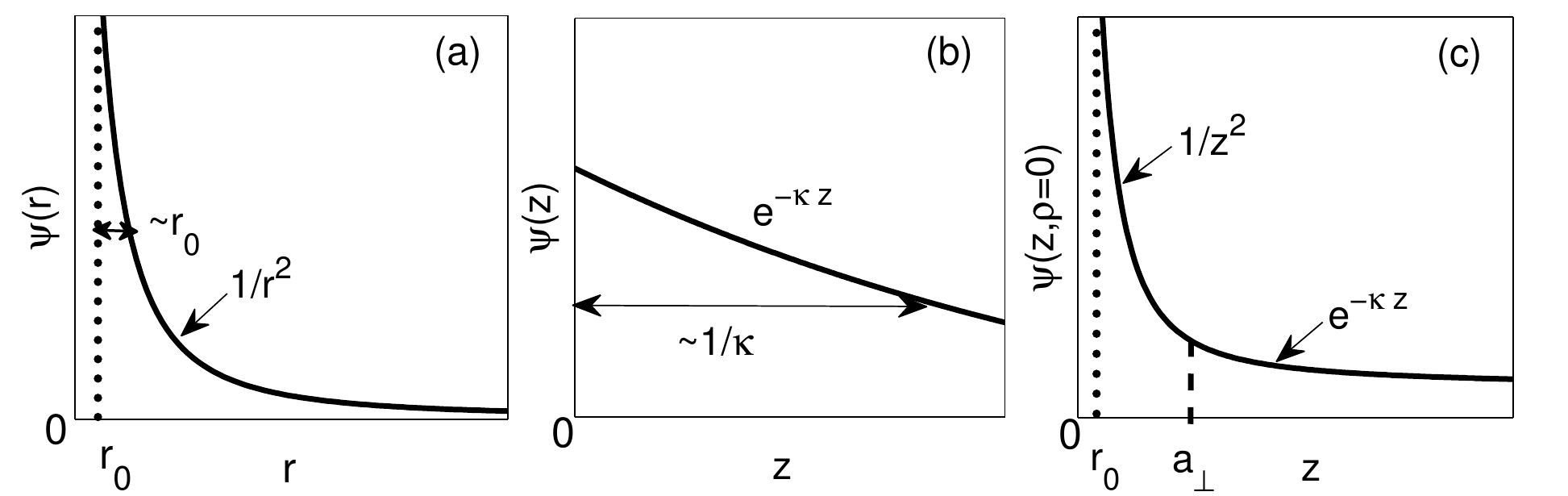}
\caption{Schematics of the wave functions for shallow molecules in different geometries: (a) the radial wave function $\psi(r)$ in 3D, (b) $\psi(z)$ in pure 1D and (c) $\psi(z,\rho=0)$ in quasi-1D. Here $r, z, \rho$ are respectively the two-body distances in 3D space, along 1D (longitudinal) and along transverse direction; $r_0$ is the short-range cutoff.  In panels (a) and (b), we show the typical lengths (horizontal lines with arrows), $l_{3D}\sim r_0$ and $l_{1D}\sim 1/\kappa$, when $\psi$ decays to half of its value at origin ($r=r_0$ or $z=0$). In panel (c), we mark the location of the transverse confinement length $a_{\perp}$, which is the typical scale separating the short-range and long-range regimes with different asymptotic behaviors.  
} \label{fig1}
\end{figure}

Physically, the strong three-body losses near the $p$-wave resonance of a 3D gas can be understood from the simple two-body bound state (molecule) property as follows. First, we write down the molecule wave function outside the interaction potential:
\begin{equation}
\Psi_{b}({\bf r})=Y_{1m}(\Omega_r) e^{-\kappa r}(\frac{1}{r^2}+\frac{\kappa}{r}),\label{psi_3d}
\end{equation}
where $m=0,\ \pm1$ is the scattering channel, ${\cp r}$ is the relative coordinate and $\kappa$ determines the binding energy $E_b=-\kappa^2/(2\mu)$ ($\mu$ is the reduced mass). For shallow molecules with $\kappa\rightarrow 0$, the wave-function simply scales as $1/r^2$ [see Fig.\ref{fig1}(a)], which is extremely singular at short distance and decays fast at long distance. In particular, it is not normalizable at $r\rightarrow 0$, and in practice one has to set a finite short-range cutoff $r_0$ to enable the normalization \cite{footnote_r0}. Consequently, the two-body wave function is highly localized in short distance [Fig.\ref{fig1}(a)] and one can easily check that most of the molecule weight lies in the short-range regime $r\gtrsim r_0$. Therefore, the molecules generated in the 3D $p$-wave resonance, even for shallow ones and open-channel-dominated ones, are very likely to decay into deep molecules due to the large wave function overlap, which cause severe atom losses in the three-body collision process. We note that this is an generic feature of high-partial-wave scattering with $l\ge 1$, where the centrifugal barrier $l(l+1)/r^2$ leads to a very singular and un-normalizable wave function, $\sim 1/r^{l+1}$, at short distance.



In this context, we pointed out in an earlier work that the 1D geometry may solve the loss problem \cite{Cui}. This is based on the following facts. Namely, in 1D there is no centrifugal barrier for $p$-wave scattering, and the two-body wave function experiences no singularity in the short-range regime except for a discontinuity due to the anti-symmetry requirement \cite{Cui}. As a result, the shallow $p$-wave molecules near the 1D resonance can be much more extended [$\sim sgn(z) e^{-\kappa |z|}$, see Fig.\ref{fig1}(b)] and thus can survive in the three-body collision into deep molecules. This makes the 1D system a promising one to suppress atom loss and meanwhile we explore the many-body physics in the presence of a resonant $p$-wave interaction. Nevertheless, we have to face the reality that in cold atom experiments all 1D systems are actually {\it quasi-1D} systems, which are generated by applying tight transverse confinements in 3D space. Such a quasi-1D system behaves as effectively 1D for long-range scattering but as 3D in the short-range regime [Fig.\ref{fig1}(c)]. 
Therefore, the $p$-wave centrifugal barrier and the wave function singularity as $1/r^2$ are still applicable to the short-range regime of two-body scattering in quasi-1D. It is thus an open question whether the pure 1D analysis can apply to the realistic quasi-1D system, which comprises the main motivation of the present work.

In this work, we study the $p$-wave scattering property of two particles in quasi-1D system. We revisit the theory of the confinement-induced $p$-wave resonance as studied in the literature \cite{CIR_p_1,CIR_p_2,CIR_p_3}, and derive two effective 1D parameters, i.e., the $p$-wave scattering length and effective range, in terms of the 3D scattering parameters and the transverse confinement length. The effective parameters are shown to well reproduce the binding energy of shallow molecules near the 1D resonance. We further calculate the wave function weight of these shallow molecules outside the short-range regime, and find that it can be much larger than the same weight of shallow molecules in 3D without confinement. This means that, by squeezing the molecules transversely via confinements, more weight moves from the (3D) short-range to the (longitudinal) long-range regime. To verify this, we show explicitly how
the molecule distribution changes as we  gradually tighten up the transverse confinement.
These results, which echo our earlier analysis of a pure 1D system \cite{Cui}, can be practically meaningful as they strongly suggest that a resonant $p$-wave system is much more stable in quasi-1D than in 3D, at least in the open-channel dominated regime.
{\it Effective scattering in quasi-1D.} We start by deriving the effective 1D parameters for $p$-wave scattering in quasi-1D. 
Given the transverse harmonic confinement with frequency $\omega_{\perp}$, the non-interaction Hamiltonian for the relative motion of two particles can be written as $H_0=-\frac{\nabla^2_{\mathbf{r}}}{2\mu}+\frac{\mu}{2}\omega_{\perp}^2 {\cp \rho}^2$, where $\rho=(x,y)$ and $\cp r=(z,\rho)$ are the relative coordinates and $\mu$ is the reduced mass. For the $p$-wave interaction with three scattering channels ($m=0,\pm 1$), we consider the typical case when the magnetic field is along the free direction ($z$). Thus only the $m=0$ channel contributes to the induced interaction along $z$. In the literature, the induced $p$-wave resonance has been studied by different methods \cite{CIR_p_1,CIR_p_2,CIR_p_3}. Here we utilize the Lippmann-Schwinger equation to write down the scattering wave function as
\begin{eqnarray}
\psi(z,\rho)
&=& \sin(k z)\phi_0(\rho)\nonumber\\
&&+f\sum_{n,q} \phi_n^*(0)\phi_n(\rho)\frac{e^{iqz}q}{E-\epsilon_n-q^2/(2\mu)+i0^+}, \label{wf}
\end{eqnarray}
where $\phi_n(\rho) (n=0,1...)$ is the eigen-state of the transverse harmonic oscillator with eigen-energy $\epsilon_n=(2n+1)\omega_{\perp}$; $E=\omega_{\perp}+k^2/(2\mu)$ is the scattering energy with $k^2\ll 2\mu\omega_{\perp}$ ensuring low-energy scattering in quasi-1D; and $f$ is a quantity related to the two-body scattering matrix, which determines both the long-range and short-range behaviors of $\psi$ as showing below.

In the long-range regime ($z\rightarrow \infty$), the wave function (\ref{wf}) is frozen at the lowest transverse mode: $\psi \rightarrow \phi_0(\rho)  \big[\sin(k z) + f_{1D}{\rm sgn}(z) e^{ik|z|} \big]$, with $f_{1D}=f\phi_0^*(0)$. Taking $f_{1D}=\sin\delta_k e^{i\delta_k}$, $\psi$ can be reduced to
\begin{equation}
\psi(z,\rho)\rightarrow \phi_0(\rho) {\rm sgn}(z) \sin(k|z|+\delta_k) , \label{wf_long}
\end{equation}
where the phase shift $\delta_k$ is related to the 1D scattering parameters via $\tan\delta_k/k=-l_p(k)$\cite{Cui}, and finally we obtain
\begin{equation}
-\frac{k}{f_{1D}}-ik=\frac{1}{l_p}-r_pk^2. \label{1d}
\end{equation}
Note that here we have used the energy-dependent scattering length following $1/l_p(k)=1/l_p-r_pk^2$, where $l_p$ is the reduced 1D scattering length and $r_p$ is the effective range.

To extract $l_p$ and $r_p$, one has to determine $f_{1D}$ by considering the short-range behavior of $\psi$. For simplicity, we take $\rho=0$ and $z\rightarrow 0^+$, and the wave function (\ref{wf}) reduces to (up to a factor $\phi_0(0)$)
\begin{equation}
\psi(z,0)= \sin(kz)+f_{1D}\left(e^{ikz}+G(z,E)\right), \label{3d_short}
\end{equation}
Here $G(z,E)=\sum_{n\ge 1}e^{-\sqrt{2\mu(\epsilon_n-E)}z} a_{\perp}^2/(2z^2)+const.+ c(E)z/a_{\perp}+o(z^2)$ as $z\rightarrow 0$, where $a_{\perp}=(\mu\omega_{\perp})^{-1/2}$ is the confinement length and the constant $c$ can be extracted as $c=-2\zeta(-1/2,1-k^2a_{\perp}^2/4)$\cite{footnote} (here $\zeta(\cdot,\cdot)$ is the Hurwitz zeta function). By matching (\ref{3d_short}) to the $p$-wave short-range boundary condition $\psi\rightarrow 1/z^2-z/(3v_p(E))$, we obtain
\begin{equation}
\frac{ka_{\perp}}{f_{1D}}+ika_{\perp}-2\zeta(-\frac{1}{2},1-\frac{k^2a_{\perp}^2}{4})=-\frac{a_{\perp}^3}{6v_p(E)}, \label{3d}
\end{equation}
where we use the energy-dependent $p$-wave scattering volume $1/v_p(E)=1/v_p-k_0/2(2\mu E)$, with $v_p$ and $k_0$ are respectively the zero-energy scattering volume and the effective range in 3D.

By combining Eqs.(\ref{1d}) and (\ref{3d}) and keeping up to the $k^2$ terms, we obtain the effective 1D parameters as
\begin{eqnarray}
\frac{1}{l_p}&=&\frac{a_{\perp}^2}{6}\left(\frac{1}{v_p}-\frac{k_0}{a_{\perp}^2}\right) -\frac{2}{a_{\perp}}\zeta(-\frac{1}{2},1);\label{l_p} \\
r_p&=&\frac{a_{\perp}^2k_0}{12}-\frac{a_{\perp}}{4}\zeta(\frac{1}{2},1). \label{r_p}
\end{eqnarray}
Equation (\ref{l_p}) predicts the confinement induced $p$-wave resonance at a critical scattering volume:
\begin{equation}
v^{(c)}_p=a_{\perp}^3\left(k_0 a_{\perp} + 12\zeta(-\frac{1}{2},1)\right)^{-1}. \label{v_pc}
\end{equation}

We note that our results of $l_p$ (\ref{l_p}) and $v_p^{(c)}$(\ref{v_pc}) are consistent with those obtained in Ref.\cite{CIR_p_3}, but differ from Refs.\cite{CIR_p_1,CIR_p_2} where the effect of zero-point energy in $E$ was not taken into account. Our result of $r_p$ (\ref{r_p}), which includes both terms from the 3D effective range and from  the renormalization of higher transverse modes, also differs from that of Ref. \cite{CIR_p_2}.

\begin{figure}[h]
\includegraphics[width=8cm]{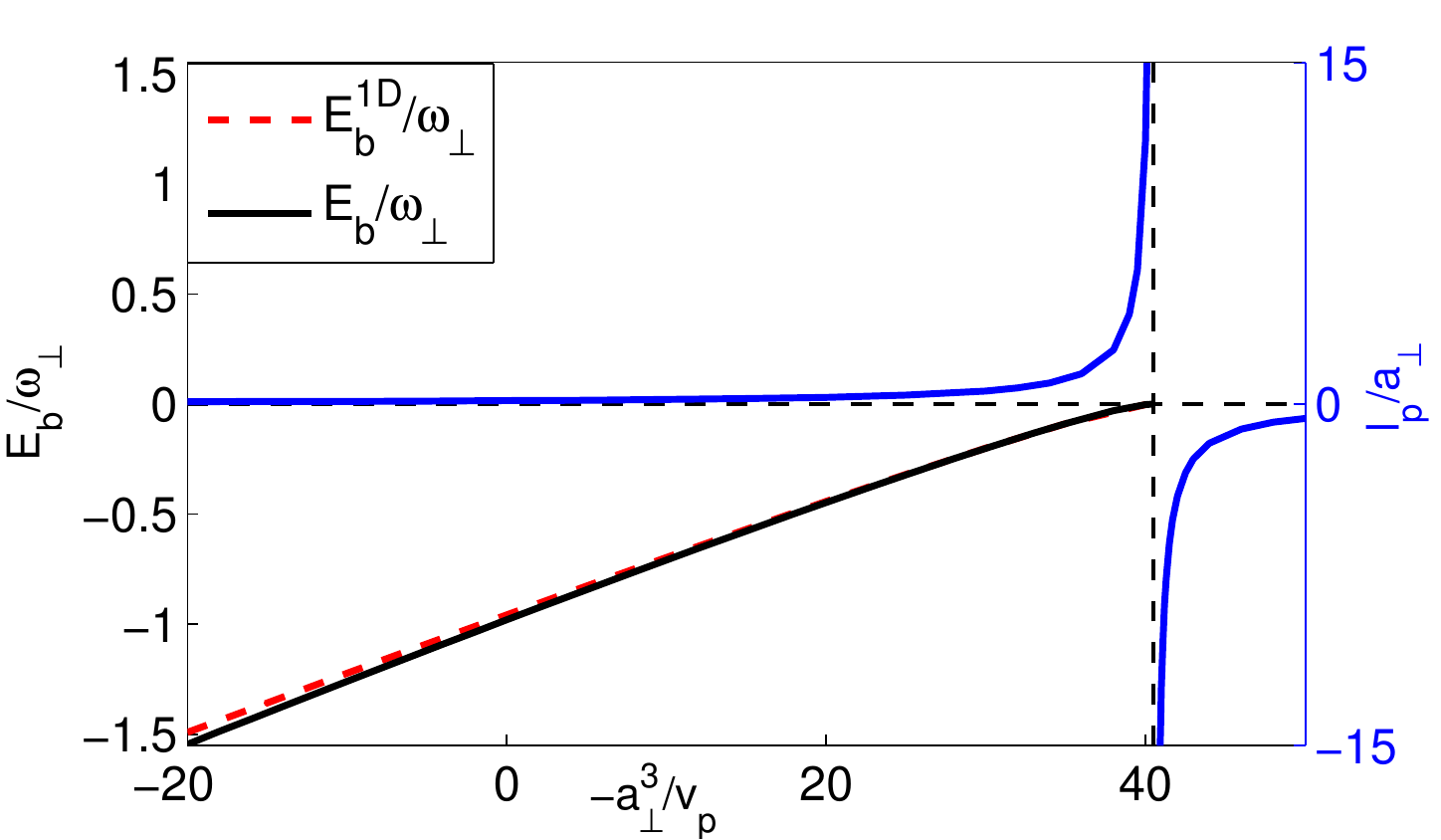}
\caption{ Effective 1D scattering length $l_p$ (blue lines) and the binding energies $E_b$(black solid) and $E_b^{1D}$ (red dashed) as functions of $-1/v_p$. Here we take a typical value of $k_0a_{\perp}=-38.0$ (see text).
The units of length and energy are respectively $a_{\perp}$ and $\omega_{\perp}$.
} \label{fig2}
\end{figure}

{\it Bound state solution.} From Eq.(\ref{3d}), we can solve the bound state with binding energy $E_b=E-\omega_{\perp}=-\kappa^2/(2\mu)<0$ through the transformations $f_{1D}\rightarrow \infty$ and $k\rightarrow i\kappa$, which gives:
\begin{equation}
-\kappa a_{\perp}-2\zeta(-\frac{1}{2},1+\frac{\kappa^2a_{\perp}^2}{4})=-\frac{a_{\perp}^3}{6v_p(E)}. \label{kappa}
\end{equation}
On the other hand, we can also use the effective 1D parameters (\ref{l_p}) and (\ref{r_p}) to determine $\kappa_{1D}$ from Eq.(\ref{1d}), which follows:
\begin{equation}
\kappa_{1D}=\frac{1}{l_p}+r_p\kappa_{1D}^2. \label{kappa_1d}
\end{equation}
Note that Eq. (\ref{kappa_1d}) can also be obtained by the methods of interaction renormalization \cite{Cui, Zinner} and boundary condition \cite{Girardeau,Blume} for 1D  $p$-wave system.

In Fig.\ref{fig2}, we show $l_p$ and the bound state solutions [$E_b=-\kappa^2/(2\mu),\ E_b^{1D}=-\kappa_{1D}^2/(2\mu)$] 
as functions of $-1/v_p$. Here we take a typical case of $^{40}$K fermions near $p$-wave resonance with $k_0=-0.04a_0^{-1}$ ($a_0$ is the Bohr radius)\cite{K40_2}, and a tight confinement length $a_{\perp}=50$nm. We can see that the induced 1D resonance ($l_p=\infty$) lies in the BCS side of the Feshbach resonance with $a_{\perp}^3/v_p^{(c)}=-40.5$. Right at 1D resonance, a $p$-wave molecule starts to emerge, and the 1D prediction $E_b^{1D}$ [from Eq.(\ref{kappa_1d})] matches well with the exact $E_b$ [from Eq.(\ref{kappa})] for shallow molecules ($-E_b\le\omega_{\perp}$). As the molecule becomes deeper, $E_b^{1D}$ starts to deviate visibly from $E_b$, which can be attributed to more and more higher traverse modes involved in the real molecule formation and thus the 1D framework breaks down. 

{\it Molecule wave function.} Following Eq.(\ref{wf}), we can write down the wave function of the bound state as:
\begin{equation}
\psi_b(z,\rho)=\frac{1}{\sqrt{\cal N}}sgn(z) \left(\sum_{n=0}^{\infty} \phi_n^*(0) \phi_n^*(\rho) e^{-\sqrt{4n/a_{\perp}^2+\kappa^2}|z|}\right),
\end{equation}
where $\cal N$ is the normalization factor. Considering the non-normalizability of $\psi$ at short-range, we have set a short-range cutoff $\pm r_0$ along $z$ for the wave function normalization:
\begin{equation}
\int d{\mathbf \rho} \int_{|z|>r_0} dz |\psi_b(z,\rho)|^2=1. \label{nor}
\end{equation}
In the following, to facilitate the comparison with 3D molecule wave function [Eq.(\ref{psi_3d}) with $m=0$], we have used the same normalization scheme [Eq.(\ref{nor})] for both cases. 


\begin{figure}[h]
\includegraphics[width=8.5cm]{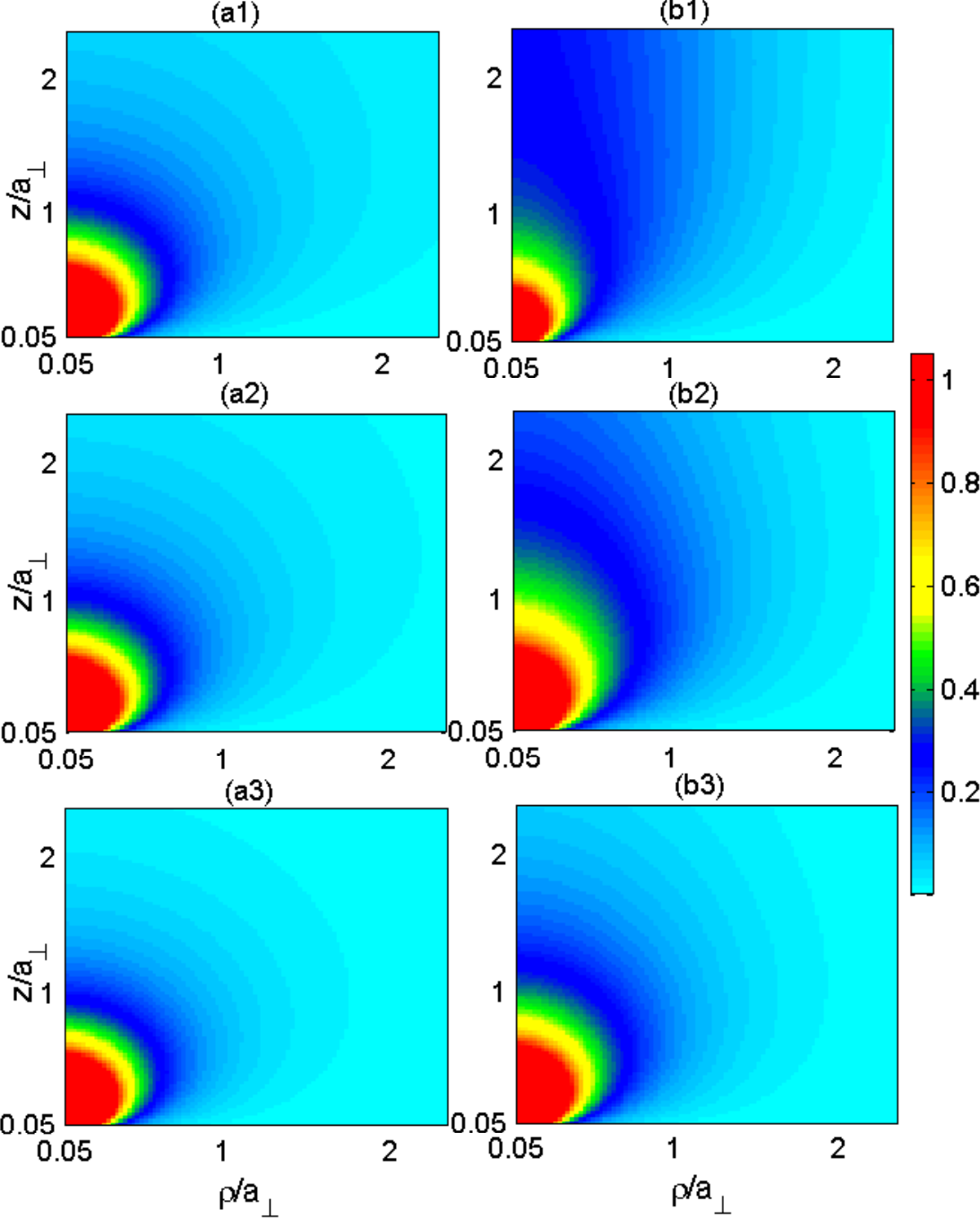}
\caption{Contour plots of the 3D (first column) and quasi-1D (second column) molecule wave functions at $\kappa a_{\perp}=0.05$ (a1,b1), $0.5$ (a2,b2) and $1$ (a3,b3). The wave functions are all normalized with a short-range cutoff $r_0=a_{\perp}/20$. 
} \label{fig3}
\end{figure}

In Fig.\ref{fig3}, we give the contour plots of the (normalized) 3D and quasi-1D wave functions in the ($z,\rho$) plane, by taking three different binding energies: $\kappa a_{\perp}=0.05,\ 0.5,\ 1$. Here we set the cutoff $r_0=a_{\perp}/20$. It is found that in general the quasi-1D molecules are more extended than 3D ones with the same binding energy, and the comparison is even more obvious for shallow molecules [see Figs.3(a1) and (b1)] with $\kappa a_{\perp}=0.05$). Therefore, we see that despite the same short-range singularity ($\sim 1/r^2$), the quasi-1D wave function in the long range regime can be significantly modified by the transverse confinement. Especially, along the longitudinal z direction, the shallow molecule in quasi-1D [Fig.\ref{fig3}(b1)] essentially follows the 1D structure and decays much more slowly than the 3D case [Fig.\ref{fig3}(a1)]. This is consistent with the schematic plots in Fig.\ref{fig1}. As the quasi-1D molecules become deeper, the wave functions become less extended along z [Fig.\ref{fig3}(b2) and (b3)] and they share more similarity with the 3D molecules [Fig.\ref{fig3}(a2)and (a3)]. In this case, the molecules lose the 1D structure and many higher transverse modes come in to take the dominated role.

\begin{figure}[h]
\includegraphics[width=8cm,height=5.5cm]{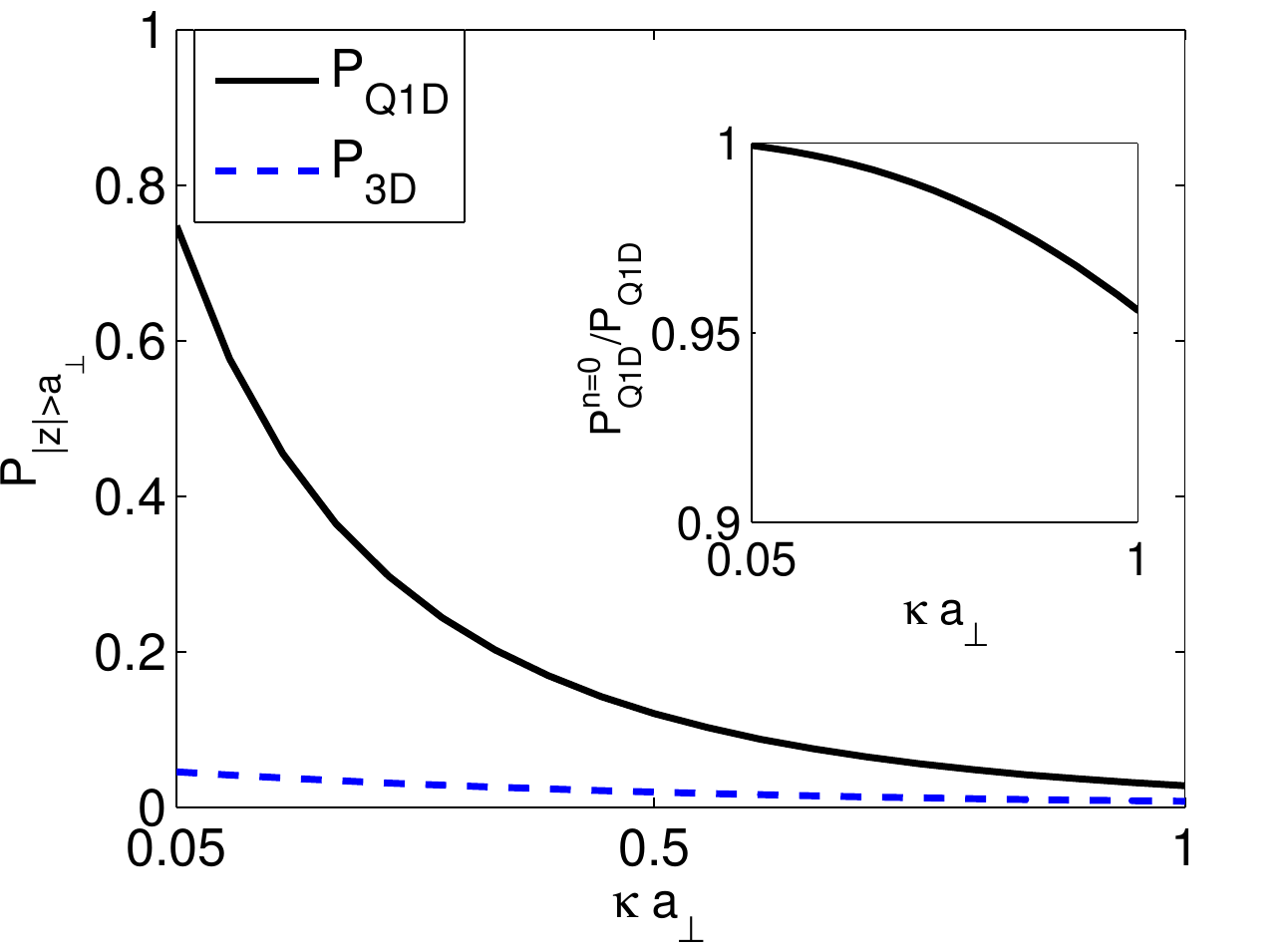}
\caption{Weights of molecules in the range $|z|>a_{\perp}$ in 3D and quasi-1D cases, denoted respectively as  $P_{3D}$ and $P_{Q1D}$, as functions of $\kappa$. Inset: the relative contribution from the lowest transverse mode $n=0$ to $P_{Q1D}$. Here $a_{\perp}=20r_0$.} \label{fig4}
\end{figure}

The different structures of 3D and quasi-1D wave functions as shown above suggest that in the latter, more weight of the molecule stay outside the short-range regime. To see this explicitly, in Fig.\ref{fig4}, we show the weights of the 3D and quasi-1D wave functions in the range $|z|>a_{\perp}$, respectively denoted as $P_{3D}$ and $P_{Q1D}$, as functions of $\kappa a_{\perp}$. We can see that for $\kappa a_{\perp}\le 0.5$, $P_{Q1D}$ is can be dozens of times larger than $P_{3D}$; for instance, when $\kappa a_{\perp}$ decreases from $0.5$ to $0.05$, $P_{Q1D}$ increases from $0.12$ to $0.75$, while $P_{3D}$ stays a small value between $0.02$ and $0.05$. Therefore the comparison is remarkable. In the inset of Fig.\ref{fig4} we show that most of the contributions to $P_{Q1D}$ are actually from the lowest transverse mode $n=0$, especially for shallow molecules. These results confirm that the quasi-1D geometry can indeed enhance the molecule occupation outside the short-range regime. 

\begin{figure}[h]
\includegraphics[width=8cm,height=5.5cm]{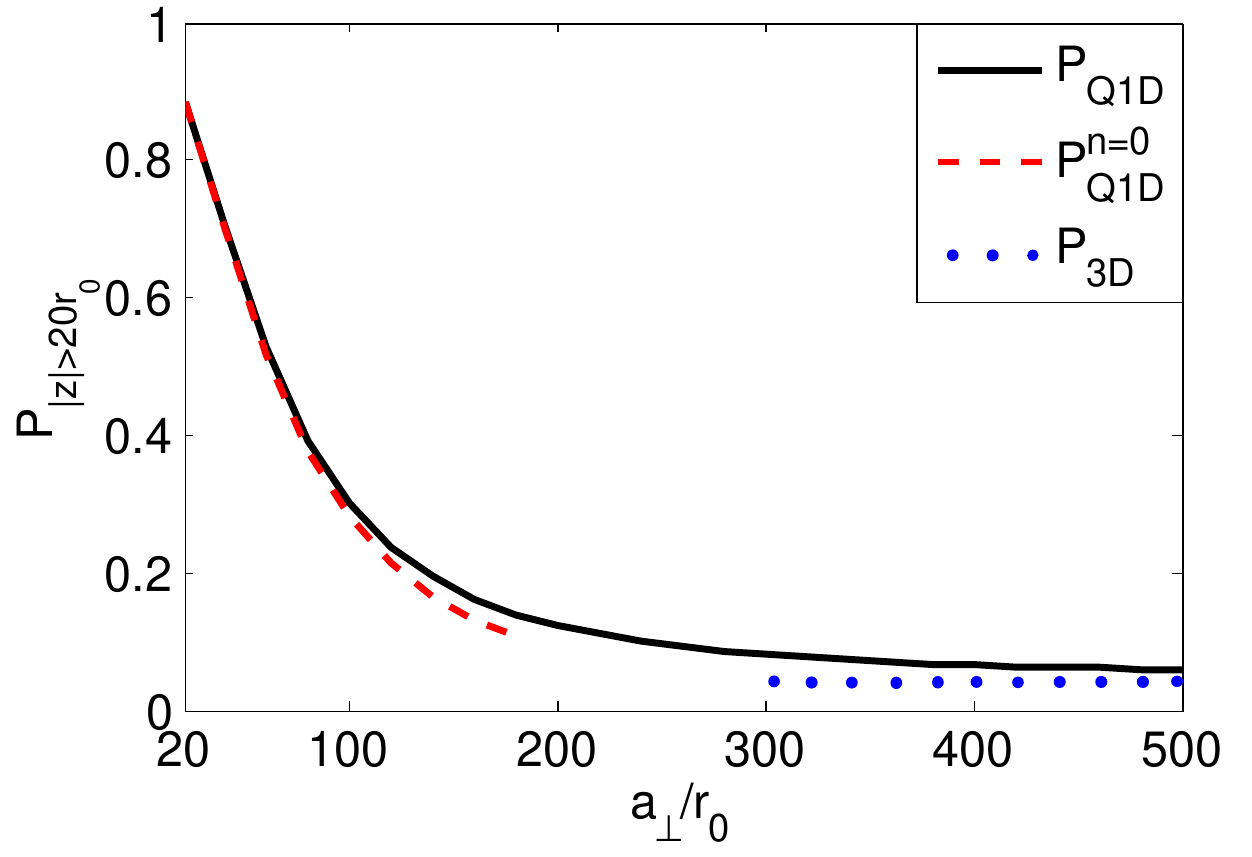}
\caption{Weight of molecules in the range $|z|>20r_0$ during the dimensional crossover from 3D to 1D ($a_{\perp}$ from $\infty$ to small). Here the length unit is $r_0$, and the binding energy is fixed by $\kappa r_0=0.001$. The red dashed line at small $a_{\perp}$ and the blue dotted line at large $a_{\perp}$ respectively show the contribution from $n=0$ mode (1D regime) and the 3D prediction (without confinement).} \label{fig5}
\end{figure}

To gain more insight into the structure of molecules changed during the 3D to 1D dimensional crossover, in Fig.\ref{fig5} we fix the molecule binding energy $\kappa r_0=0.001$ and show the molecule weight at $|z|>20r_0$ (denoted by $P_{Q1D}$) as a function of $a_{\perp}$. We can see that, as $a_{\perp}$ decreases from $\infty$ to $20r_0$, which corresponds to the system gradually evolving from 3D to quasi-1D, $P_{Q1D}$ gradually increases from a very small value ($\sim 0.05$) to a quite large one near unity ($\sim 0.9$). This shows that during the dimensional crossover from 3D to quasi-1D, the shallow molecules gradually accumulate more weight in the long-range regime. Combining with Fig.\ref{fig4}, a physical picture is that, by gradually squeezing the molecule through transverse confinements, more weight moves from the short-range to the (longitudinal) long-range regime and the molecule is much more stretched in spatial space (see also Fig.\ref{fig4}). In the extremely squeezed regime under tight confinements, the quasi-1D molecules are essentially contributed from the lowest transverse mode (red dashed line in Fig.\ref{fig5}). In this limit our previous analysis based on a pure 1D system can be automatically recovered\cite{Cui}.


{\it Summary and discussion.} In this work, we show that the shallow $p$-wave molecules in a quasi-1D system can be well predicted by the effective 1D parameters shown in Eqs.(\ref{l_p}) and (\ref{r_p}). Importantly, these shallow molecules hold most of the weight outside the short-range regime, in contrast to those in 3D without confinements. These results may serve as a guideline for detecting shallow $p$-wave molecules in quasi-1D atomic gas, which has not been achieved so far.

Meanwhile, our results strongly indicate that the $p$-wave interacting atomic gas can be much more stable against three-body loss near the resonance in quasi-1D, as compared to 3D case, especially when the $p$-wave molecules are open-channel dominated.  On the other hand, there could be another loss mechanism, such as the two-body ones due to the coupling between the relative and center-of-mass motions under the lattice confinement \cite{split_res} or due to the relaxation to lower hyperfine state \cite{K40, Gora}. So it still remains to see how the actual loss behaves in a realistic experiment. We note that a previous experiment detected the $p$-wave atom loss with transverse confinement \cite{CIR_p_expe}, but that was not in the effective 1D regime \cite{footnote2}.
We hope this work can stimulate more studies on the exciting field of 1D $p$-wave gas, which can be a natural platform to host Majorana fermions in lattices \cite{Cui2}.


{\it Acknowledgments.}  We thank Tin-Lun Ho, Meng Khoon Tey, and Nikolaj T. Zinner for helpful comments on the manuscript. The work is supported by the National Natural Science Foundation of China (No.11622436, No.11374177, No.11421092, No.11534014).


\begin{thebibliography}{99}

\bibitem{Gurarie}V. Gurarie, L. Radzihovsky, and A.V. Andreev, Phys. Rev. Lett.  {\bf 94}, 230403 (2005).

\bibitem{Yip} C.-H. Cheng and S.-K. Yip, Phys. Rev. Lett. {\bf 95}, 070404 (2005).

\bibitem{Read} N. Read and D. Greene, Phys. Rev. B {\bf 61}, 10267 (2000).

\bibitem{Zoller} S. Tewari, S. Das Sarma, C. Nayak, C. Zhang and P. Zoller, Phys. Rev. Lett. {\bf 98}, 010506 (2007).

\bibitem{Kitaev}A.Y. Kitaev, Phys. Usp. {\bf 44}, 131 (2001).


\bibitem{K40} C. A. Regal, C. Ticknor, J. L. Bohn, and D. S. Jin, Phys. Rev. Lett. {\bf 90}, 053201 (2003).

\bibitem{K40_2}C. Ticknor, C. A. Regal, D. S. Jin, and J. L. Bohn, Phys. Rev. A {\bf 69}, 042712 (2004).

\bibitem{Toronto}
C. Luciuk, S. Trotzky, S. Smale, Z. Yu, S. Zhang, J. H. Thywissen, Nature Physics {\bf 12}, 599 (2016).

\bibitem{Li6_1}J. Zhang, E. G. M. van Kempen, T. Bourdel, L. Khaykovich, J. Cubizolles, F. Chevy, M. Teichmann, L. Tarruell, S. J. J. M. F. Kokkelmans, and C. Salomon, Phys. Rev. A {\bf 70}, 030702 (R)(2004).

\bibitem{Li6_2}
C. H. Schunck, M. W. Zwierlein, C. A. Stan, S. M. F. Raupach, W. Ketterle, A. Simoni, E. Tiesinga, C. J. Williams, and P. S. Julienne, Phys. Rev. A {\bf 71}, 045601 (2005).

\bibitem{Rb}S. Dong, Y. Cui, C. Shen, Y. Wu, M. K. Tey, L. You and B. Gao, Phys. Rev. A {\bf 94}, 062702 (2016).

\bibitem{private}M. K. Tey (Private communication).

\bibitem{footnote_r0} In general $r_0$ sets the range of the interaction potential, which is typically of  the order of the Van der Waals length in realistic atomic systems.

\bibitem{Cui} X. Cui, Phys. Rev. A {\bf 94}, 043636 (2016); The two-channel generalization is given in X. Cui and H. Dong, $ibid$. {\bf 94}, 063650 (2016).

\bibitem{CIR_p_1}B. E. Granger and D. Blume, Phys. Rev. Lett. {\bf 92}, 133202
(2004).
\bibitem{CIR_p_2}L. Pricoupenko, Phys. Rev. Lett. {\bf 100}, 170404 (2008).
\bibitem{CIR_p_3}S.-G. Peng, S. Tan, and K. Jiang, Phys. Rev. Lett. {\bf 112}, 250401 (2014).
\bibitem{footnote}By using the relation $\int_0^{\infty} e^{-2\sqrt{n}|z|/a_{\perp}}dn=a_{\perp}^2/(2z^2)$, $c$ \\can  be obtained as $c=-2 \lim_{\Lambda\rightarrow \infty}( \sum_{n=0}^{\Lambda} \sqrt{n+\alpha}-$\\$2/3 (\Lambda+\alpha)^{3/2} )=-2\zeta(-1/2,\alpha)$, with $\alpha=1-k^2a_{\perp}^2/4$.

\bibitem{Zinner} $p$-wave renormalization was also discussed in the 1D lattice system: M. Valiente, N. T. Zinner, Few-Body Systems {\bf 56}, 845 (2015).

\bibitem{Girardeau} M. D. Girardeau, and M. Olshanii, Phys. Rev. A {\bf 70}, 023608 (2004).

\bibitem{Blume}K. Kanjilal and D. Blume, Phys. Rev. A, {\bf 70}, 042709 (2004).




\bibitem{split_res} S. Sala, P.-I. Schneider, and A. Saenz, Phys. Rev. Lett. {\bf 109}, 073201 (2012); S. Sala, G. Z{\" u}rn, T. Lompe, A.N. Wenz, S. Murmann, F. Serwane, S. Jochim and A. Saenz, $ibid$. {\bf 110}, 203202 (2013); 

\bibitem{Gora}D.V. Kurlov, G.V. Shlyapnikov, Phys. Rev. A {\bf 95}, 032710 (2017).

\bibitem{CIR_p_expe} K. Gunter, T. Stoferle, H. Moritz, M. Kohl, and T. Esslinger, Phys. Rev. Lett. {\bf 95}, 230401 (2005).

\bibitem{footnote2} The atom number in the experiment\cite{CIR_p_expe} is much larger than the aspect ratio of trapping potential, so a large portion of the atoms occupies in the higher transverse modes.

\bibitem{Cui2} X. Cui, Phys. Rev. A {\bf 95}, 041601(R) (2017).

\end{thebibliography}
\end{document}